# Physics Archives

*November 2009*

*Brainstorming through the Sequence Universe:*
*Theories on the Protein Problem*

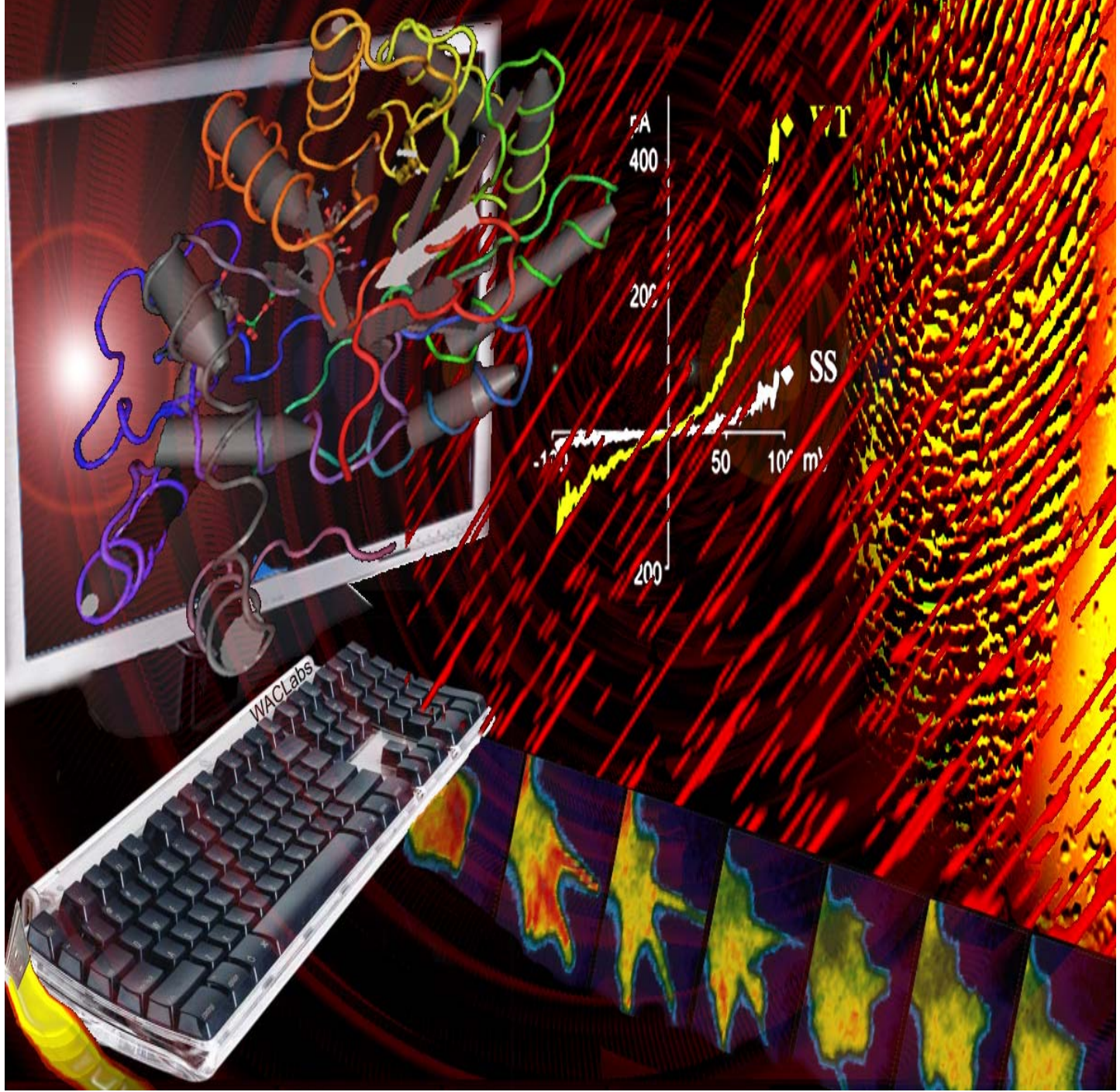

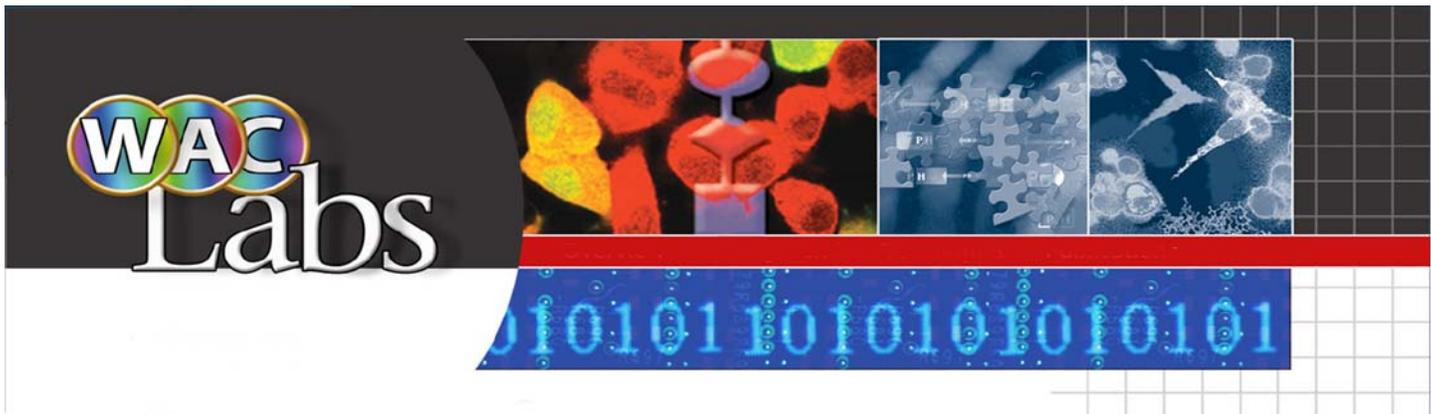

# Brainstorming through the Sequence Universe: Theories on the Protein Problem

Kyung Dae Ko[1,2,*], Yoojin Hong[1,3,*], Gaurav Bhardwaj[1,2], Teresa M. Killick[1,2], Damian B. van Rossum[1,2,¥], Randen L. Patterson[1,2,¥]

(1) Center for Computational Proteomics, The Pennsylvania State University
(2) Department of Biology, The Pennsylvania State University
(3) Department of Computer Science, The Pennsylvania State University

[*] These authors contributed equally to this work
[¥] Address correspondence to:

Randen L. Patterson, 230 Life Science Bldg, University Park, PA 16802. Tel: 001-814-865-1668; Fax: 001-814-863-1357; E-mail: rlp25@psu.edu.

Damian B. van Rossum, 518 Wartik Laboratory, University Park, PA 16802. Tel: 001-814-863-1007; Fax: 001-814-863-1357; E-mail: dbv10@psu.edu.

**Abstract**
     **Just as physicists strive to develop a TOE (theory of everything), which explains and unifies the physical laws of the universe, the life-scientist wishes to uncover the TOE as it relates to cellular systems. This can only be achieved with a quantitative platform that can comprehensively deduce and relate protein structure, functional, and evolution of genomes and proteomes in a comparative fashion. Were this perfected, proper analyses would start to uncover the underlying physical laws governing the emergent behavior of biological systems and the evolutionary pressures responsible for functional innovation. In the near term, such methodology would allow the vast quantities of uncharacterized (e.g. metagenomic samples) primary amino acid sequences to be rapidly decoded. Analogous to natural products found in the Amazon, genomes of living organisms contain large numbers of proteins that would prove useful as new therapeutics for human health, energy sources, and/or waste management solutions if they could be identified and characterized. We previously theorized that phylogenetic profiles could provide a quantitative platform for obtaining unified measures of structure, function, and evolution (SF&E)(1). In the present manuscript, we present data that support this theory and demonstrates how refinements of our analysis algorithms improve the performance of phylogenetic profiles for deriving structural/functional relationships.**

**Introduction**

   The 'protein problem' has remained unsolved despite decades of research(2;3). In principle, one expects that the primary amino acid sequence of a protein determines its structure, function, and evolutionary (SF&E) characteristics. Yet, there still is no reliable method for predicting the native state structure of a protein and its function given only its sequence. Although the number of putative protein sequences of a given length is enormous, very few of these would fold rapidly and reproducibly and have useful function. In addition, inferring the evolutionary relationships among highly divergent protein sequences is a daunting task.  In general, when pairwise sequence alignments between protein sequences fall below 25% identity, statistical measurements do not provide support robust enough to identify clear phylogenetic relationships, structural features, or protein function despite intensive research in this area(2;4;5).  In this manuscript, we propose our updated strategies towards solving the 'protein problem' which are: (i) simple yet powerful, (ii) broad and relate to diverse aspects of proteins, (iii) interdisciplinary, and (iv) significant for the understanding of life.

   The anchor for our approach lies in the discovery that phylogenetic profiles are capable of measuring the fundamental properties of proteins.  Through construction of knowledge-base profiles related for either structural or functional qualities, phylogenetic profiles provide measurements which can be used to simultaneously infer evolutionary distances, create functional models, and identify structural components of protein sequences from any organism(6-8).  Indeed, our recent publications and ongoing research demonstrate that SF&E information obtained from phylogenetic profiles are exceptionally informative to our laboratory experiments at multiple scales (e.g. whole protein, single protein domain, and single amino acid)(1;9-13).  We have used these analyses: (i) to reconstruct evolutionary histories(1;9), (ii) to identify functions in domains of unknown function(1;11;12;14-16), (iii) to classify structural homologues of high sequence divergence(1)(Hong et al, Physics Archives November 2009), and (iv) to inform our biochemical experimentation by isolating key amino acids important to protein function(10-12;15).

   The power behind our phylogenetic profiles is derived from the Position-Specific Scoring Matrices (PSSMs, i.e. profile), which contain a frequency table for substitutions that occur in related sequences; PSSMs are a powerful measure of homology(17-19).  Indeed, it is well-established that PSSMs contain more information than individual sequences.  We take advantage of the increased information content of PSSMs and quantify their alignments within a phylogenetic profile.  Under this paradigm, a protein is defined as a vector where each entry quantifies the alignments of a query sequence with a PSSM.

   Our strategy is designed to amplify and encode the pair-wise alignments possible for any given query sequence in a standardized manner (see Figure 1 and Hong et al Physics Archives 2009 for full descriptions).  The output of these comparisons is a composite [product] score of either zero [when there is no significant match] or a positive value [which measures the degree of successful matching of the protein sequence to each of the PSSMs]. This procedure can be readily adapted to make an unbiased comparison between a series of query sequences by subjecting them to the same screening analysis with the same set of PSSMs. Thus, each query sequence is represented in a vector of non-negative numbers in M dimensions (M= # of PSSMs tested).  Importantly, M dimensions can also be derived from standardized measures of structural and chemical features of the query sequence.  Further power in using phylogenetic profiles as a framework for data encoding comes from the interoperability of the data matrix itself.  For example, phylogenetic profile data spaces can be analyzed using a variety of distance measurements, clustering algorithms, and information visualization algorithms, to name a few.  We propose that it is this ability to compare the results obtained from multiple algorithms using the same NXM data space that empowers protein-based phylogenetic profiles.

Our strategy is unlike most other computational approaches which generally focus on structure, function, or evolution individually. Based on our recent advances, we have determined that phylogenetic profiles provide a robust and unified framework to decode SF&E information of protein sequences. Using this paradigm we theorize that: (i) functional and evolutionary measurements can quantitatively inform structural modeling to derive accurate atomic resolution structures; (ii) structural and functional measurements can inform evolutionary histories to derive accurate evolutionary rates, deep-branch relationships, and homologous regions within each protein; and (iii) structural and evolutionary measures can provide information as to the location of functionalities and/or regulatory-elements contained within any protein. Should this theory be true, the speed at which biological processes and signaling networks would be decoded would be significantly increased. In addition, it would drastically improve our ability to pinpoint functional targets for currently intractable problems such as developing mutation resistant pharmacophore therapeutics, predicting functional mutations that may allow for cross-species viral infections, and identifying the functional consequences of disease-causing polymorphisms in humans. Perhaps most importantly, phylogenetic profiles perform in the "twilight zone" of sequence similarity(1;9). Therefore, this approach can be harnessed to decode the most challenging protein datasets, and be scaled up to screen proteomes and the vast quantities of sequences being obtained from metagenomic and other large scale sequencing projects. In the present manuscript, we report our recent advances in deriving useful measurements using phylogenetic profiles and improving our analysis of these datasets.

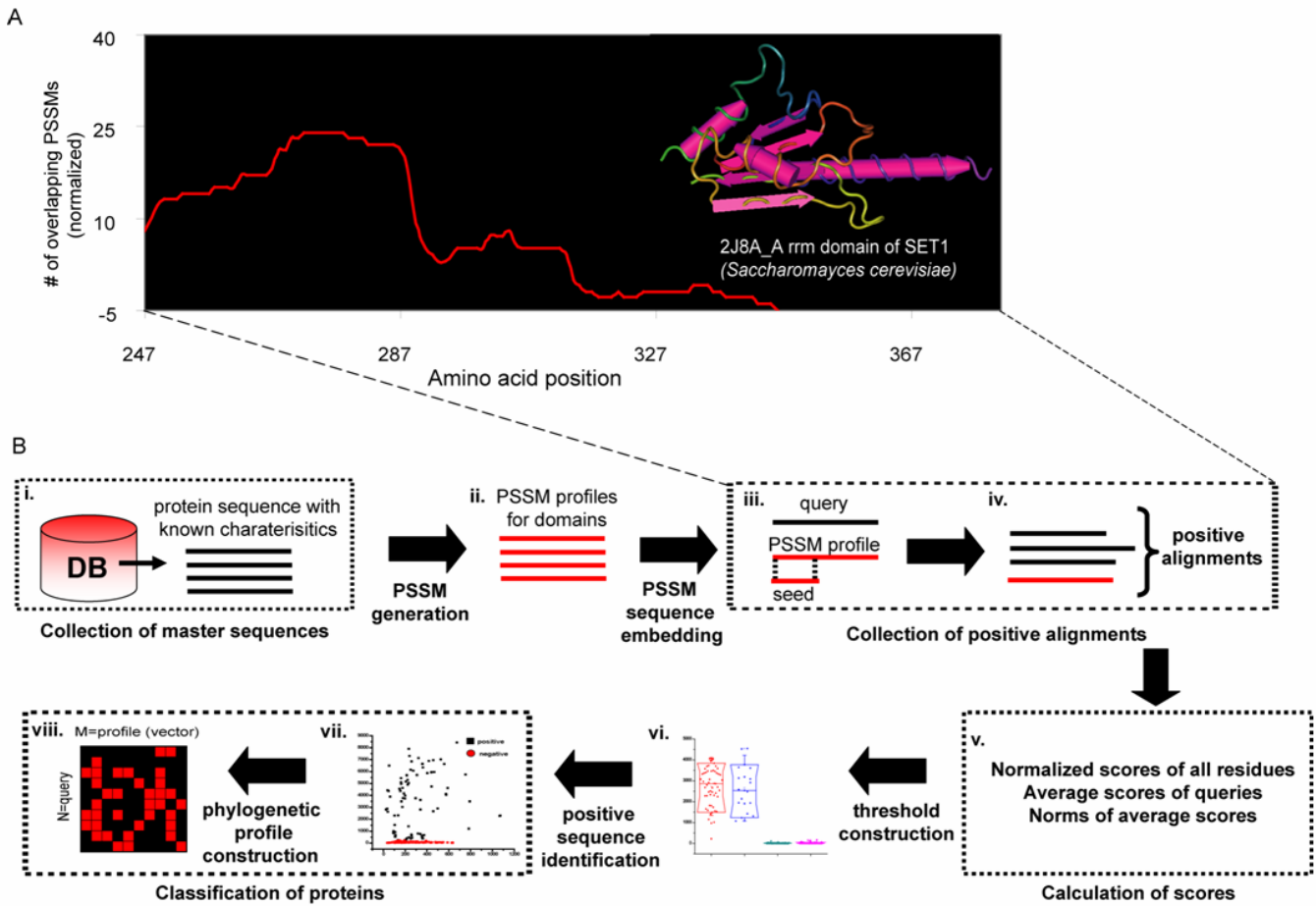

Figure 1. Phylogenetic Profile Generation- (A) Histogram displaying the position of the alignments obtained from GDDA-BLAST with 100 RRM-specific PSSMs; *inset* the X-ray structure of the N-terminus RRM domain of SET1. (B) The schema of the classification method; (i) master sequences are collected from databases such as CDD and PDB, (ii) PSSMs are then generated from the master sequences, (iii) PSSM sequences are then embedded into every position of the query sequence, (iv) query sequences are then aligned with the PSSM library. Positive alignments above threshold (60% coverage, 10% identity) are collected, (v) scores are calculated, (vi) thresholds are constructed, (vii) sequences scoring above threshold are collected, (viii) phylogenetic profiles are constructed (see Methods for full description).

# Results
## *Theories on Structure/Function*

One of the major challenges in the genomic era is annotating function to the vast quantities of sequence information now available.  Indeed, most of the protein sequence database lacks comprehensive annotation, even when experimental evidence exists.  Further, within structurally resolved and functionally annotated protein domains, additional functionalities contained in these domains are not apparent.  To add further complication, small changes in the amino-acid sequence can lead to profound changes in both structure and function, underscoring the need for rapid and reliable methods to analyze these types of data.  Indeed, most proteins when queried using popular algorithms for identifying protein function (e.g. NCBI CDD, Pfam, SMART, InterProscan) return few, if any, results that are statistically significant(20-23).  In an attempt to overcome these limitations, we have utilized phylogenetic profiles, and have had some remarkable results.  We have been successful in identifying fundamental protein interactions (e.g. protein-nucleic acid, protein-lipid, protein-protein, protein-small molecule)(1;10-15).  Throughout this work, we have made a number of predictions such as: (i) PSSMs of related function can contain "homology" even if they represent different structures, (ii) PSSMs libraries generated using specific folds can accurately identify homologous folds, (iii) PSSM libraries generated for a specific activity can accurately identify homologous functions in proteins of diverse structure, as well as differentiating activity within a specific fold, and (iv)  single-amino acid phylogenetic profiles can obtain quantitative functional thresholds.

These theories are supported by our analysis of RNA-binding proteins (RBPs).  RNAs in a cell generally have many functions such as (i) a carrier of genetic information, (ii) a catalyst of biochemical reactions, (iii) an adapter molecule in protein synthesis, and (iv) a regulator of RNA splicing and maintenance of the telomeres or chromosome ends. If we are to identify the all the functions of any specific RNA, we need to understand the functions of their binding proteins as they can control post-transcriptional processes such as pre-mRNA processing, splicing, and translation, and likely regulate RNA-enzymatic activity (e.g. dicer).  However, Since RNA structures are varied; the structures of RBPs that interact with RNA are diverse. Indeed, the known RBPs can be classified into six families by their basic binding motifs (24), and as a group RBPs have wildly different structures. In fact, even within the same RBP family, the RNA interaction site within the structure need not be conserved (24). To compound these problems, not all members of a particular fold-specific group bind to RNA. Taken together, it is difficult to identify RBPs either computationally or biochemically.

Multiple computational algorithms have been developed to attack these questions. Examples include support vector machine learning (SVM) and Hidden Markov Models (HMM) (25;26). Although these methods have utility, they suffer in divergent datasets, as well as in correctly characterizing the nucleotide substrate (e.g. DNA vs RNA).  Based on our previous studies on lipid-binding, it is reasonable to consider that specific PSSM libraries can be constructed for these attributes. To test this hypothesis, we constructed RBP-specific PSSM libraries and quantified the performance of phylogenetic profiles for the characterization of multiple benchmark data sets of RBPs (see Methods for complete description).  In total, our results provide substantial evidence that PSSM libraries can be used to obtain quantitative thresholds for RBPs and other nucleic acid-protein interactions. Further, we propose that this strategy can be implemented to obtain refined measurements for any protein function.

## *Structural/Functional Data*

Our first training set consists of 55 RNA-recognition Motifs (RRM) domain-containing sequences and negative sequences; all of which were curated from the yeast database and the Protein Databank (PDB). Generally, RRM domains are ~90 amino acids long and are composed of four beta-sheets packed against two alpha-helices. These motifs are often found in tandem within protein sequences and are known to bind a variable number (2-8) of RNA molecules (27). In addition, this motif also appears in a few single stranded DNA-binding proteins, and can also facilitate protein-protein interactions instead of nucleotides (24;28). Due to the vast literature on this class of proteins, RRMs provide an excellent model system for testing our hypotheses. Importantly, all of the negative sequences were biochemically demonstrated not to bind nucleic acids (29). All datasets, raw data, FASTA files for sequences and scripts used in the manuscript are available upon request.

Figure 1 depicts the strategy we employed. Initially, we collect master sequences from a biological database. In this case, master sequences were identified from the non-redundant NCBI database using PSI-BLAST queried with known RRM domains (e-value ≤ $1 \times 10^{-2}$). Next, we generate PSSM profiles from these master sequences (see Methods). Following, query sequences are aligned using GDDA-BLAST as previously described (1;9). In brief, query sequences are modified by adding a standard sequence length from a PSSM ("seed"), which creates a consistent initiation site for alignments (see Methods for complete description). "Seeds" in this study are generated from profiles by taking a 10% of the profile sequence from the N-terminus and C-terminus of the PSSM. Query sequences are then aligned with the parent PSSMs using reverse position specific-BLAST (rps-BLAST). From these data, % identity and % coverage are used for alignment filtering. Alignments above threshold are used to calculate normalized scores for all residues in the query sequence, the average scores of each query, and norms of the average scores on the basis of the positive alignments (for a complete description see Methods). Following, false positive sequences can be filtered by thresholds derived from the norms (average scores) for all queries. All scripts, PSSM libraries, and sequence FASTA files used in this study are available upon request.

### *Characterizing single-fold RBPs:*

Figure 1A (*inset*) depicts the structure of an atypical (truncated) RRM domain from yeast SET1 protein, a histone methyltransferase. Popular domain-prediction algorithms such as NCBI Conserved Domain Database (CDD) and Interproscan do not detect this RRM domain within SET1 at statistically significant thresholds (data not shown). When we analyzed SET1 using Gestalt Domain Detection Algorithm Basic Local Alignment Tool (GDDA-BLAST) to generate phylogenetic profiles (1;9) with 100 RRM-specific PSSMs, we observe that these profiles align in SET1 and correspond to the position of the RRM domain (Figure 1A, see Methods). Thus, these PSSM profiles are sensitive enough to detect this highly divergent RRM domain.

Figure 2A shows that the sequences in the training dataset are completely separated into positive and negative groups. The minimum score of the positive group is ~224 and the maximum score for the negative group is ~127. Using the same library of PSSMs, we then performed the same analysis on a testing set. The testing set consists of 20 RRM containing sequences and 135 negative sequences. We achieve 100% accuracy in this testing set using the threshold derived from our training set (Figure 2B-C). We compared our performance to two other popular algorithms (Interproscan and SVM-PROT) (Figure 3A-B)(23;30). We observe that phylogenetic profiles and Interproscan provide robust measures, while SVM-PROT does not perform well in either the training or the testing dataset.

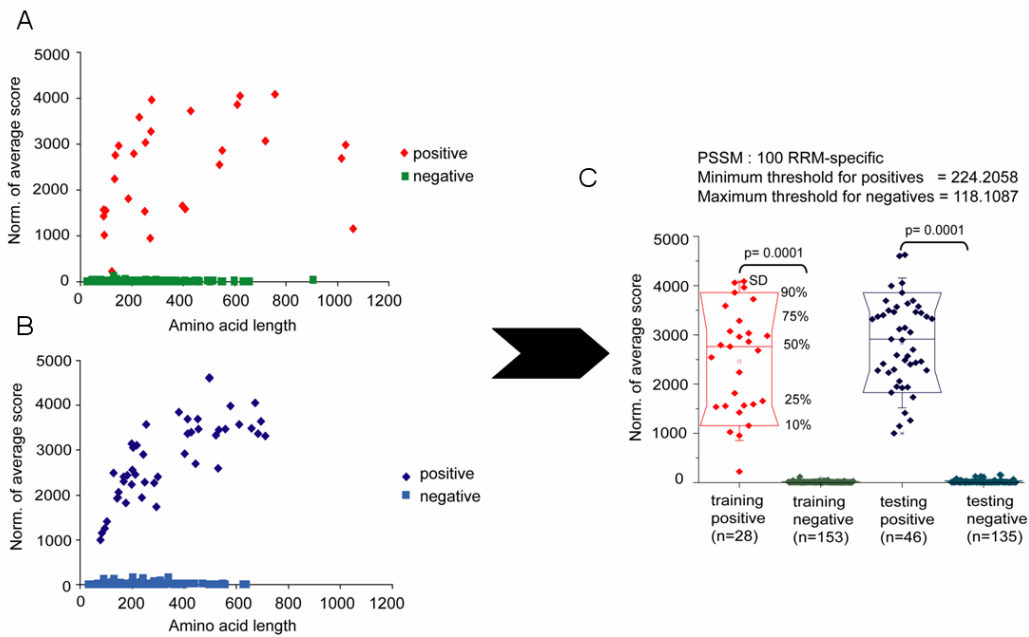

**Figure 2. Classifying RRM domains-** **(A)** Norm of average scores for 29 positive (red) and 153 negative (green) sequences when analyzed with 100 RRM-domain specific PSSMs. In this training set, the minimum value obtained for positive sequences is 224.2058. **(B)** Norm of average scores for 46 positive (dark blue) and 135 negative (light blue) sequences when analyzed with 100 RRM-domain specific PSSMs. **(C)** Box-plots with confidence values for all data in A and B. Using the threshold obtained from our training set, we achieve 100% accuracy in our testing set at significant values.

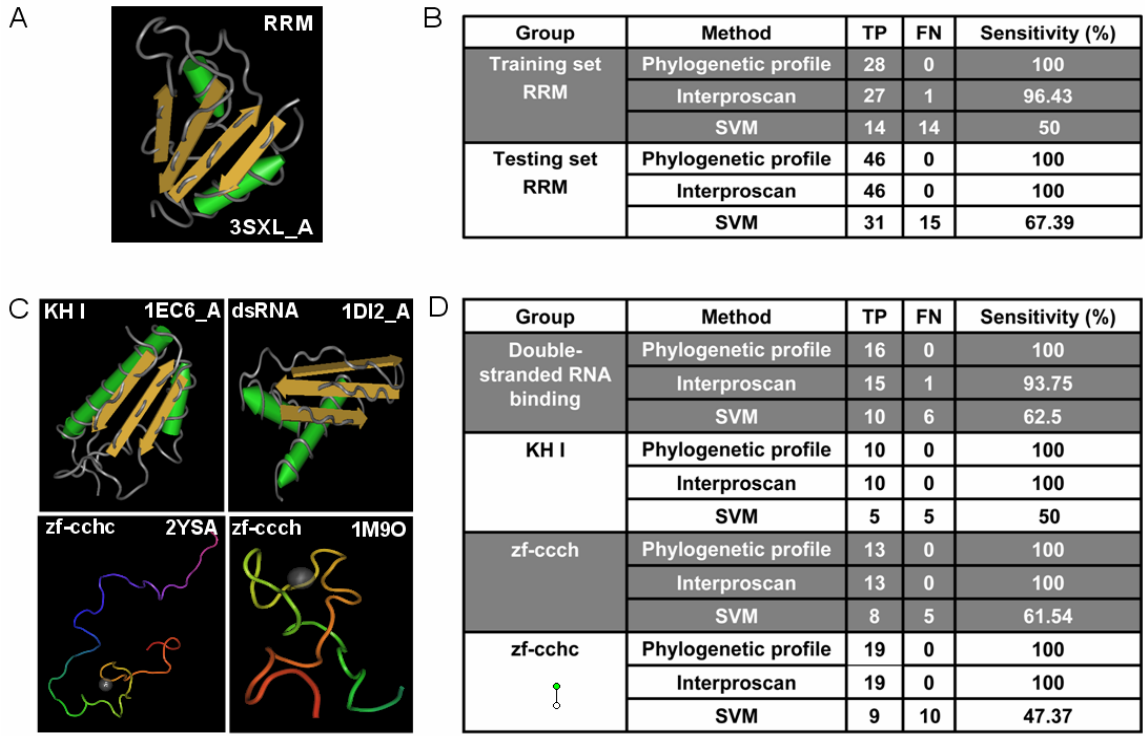

Figure 3. Single-fold RBP comparisons- Table displaying the results obtained from phylogenetic profiles, Interproscan, and support-vector machine learning (SVM-PROT) for the five RNA binding protein folds. (A) RRM structure. (B) Results for RRM training and testing datasets. (C) KH1, ssRNA-binding, zf-ccch, and zf-cchc structures. (D) Results for KH1, ssRNA-binding, zf-ccch, and zf-cchc testing datasets. All raw data, testing and training sequences used are available upon request.

To extend upon these findings, we performed similar analyses for additional classes of single-fold RBPs. These include KH1, double-stranded RNA, and zinc fingers (24). The results from these experiments are provided in Figure 3C-D. We observe that phylogenetic profiles have 100% accuracy for all single-fold RBPs tested. In comparison, Interproscan performs similarly to phylogenetic profiles while SVM-PROT performs poorly in all of these datasets.

### Characterizing RBPs containing multiple folds:

In order to comprehensively catalogue RBPs, methods are needed that can account for both characterized and uncharacterized folds. Four major classes of RBPs contain multiple folds for their respective target (24). These include mRNA, tRNA, rRNA, and snRNA. For these classes, we created PSSM libraries for the diverse folds that can accommodate these interactions. The results from our testing set are presented in Figure 4. We achieve 100% accuracy in these datasets. Comparatively, SVM-PROT performs poorly in all data sets. We could not compare the performance of Interproscan for predicting RNA-binding function as it does not model the functionality *per se*; rather it provides fold-specific information.

| RBP Group | Method | TP | FN | Sensitivity (%) |
|---|---|---|---|---|
| mRNA binding | Phylogenetic profile | 11 | 0 | 100 |
| | SVM | 4 | 7 | 36.36 |
| tRNA binding | Phylogenetic profile | 13 | 0 | 100 |
| | SVM | 10 | 3 | 76.92 |
| rRNA binding | Phylogenetic profile | 11 | 0 | 100 |
| | SVM | 7 | 4 | 63.64 |
| snRNA binding | Phylogenetic profile | 11 | 0 | 100 |
| | SVM | 3 | 8 | 27.28 |

Figure 4. Multi-fold RBP comparisons- Table comparing the results between phylogenetic profiles and SVM algorithms for four multi-fold RBP training sets (all sequences used for training, PSSMs in each library, and raw data are available upon request). We obtained thresholds of 172.4688 for mRNA binding proteins (11 sequences, 332 PSSMs), 444.5183 for tRNA binding proteins (13 sequences, 313 PSSMs), 208.3474 for rRNA binding proteins (11 sequences, 632 PSSMs), and 406.449 for snRNA binding proteins (11 sequences, 208 PSSMs).

In 2008, Shazman *et al.* demonstrated that SVM methods could be improved by incorporating electrostatic surface patch information into their analyses (31). Hence, this RBP study provides an excellent benchmark dataset, as well as another algorithm with which to compare our performance. Initially, our preliminary experiment using this testing set was performed with fold-specific PSSMs; however, these fold-specific PSSMs were insensitive in this dataset. We wondered whether our results could be improved by merely increasing the number of PSSMs in our library. To accomplish this task, we searched the PROSITE database for the key-word "RNA-binding" (32). The results from this search were then manually confirmed to ensure the specificity of these sequences. Importantly, the structure of these sequences was not taken into account for inclusion in the PSSM library. Following, additional sequences were identified by statistically significant sequence similarity and PSSMs were generated from the non-redundant NCBI database using PSI-BLAST (see Methods)(17).

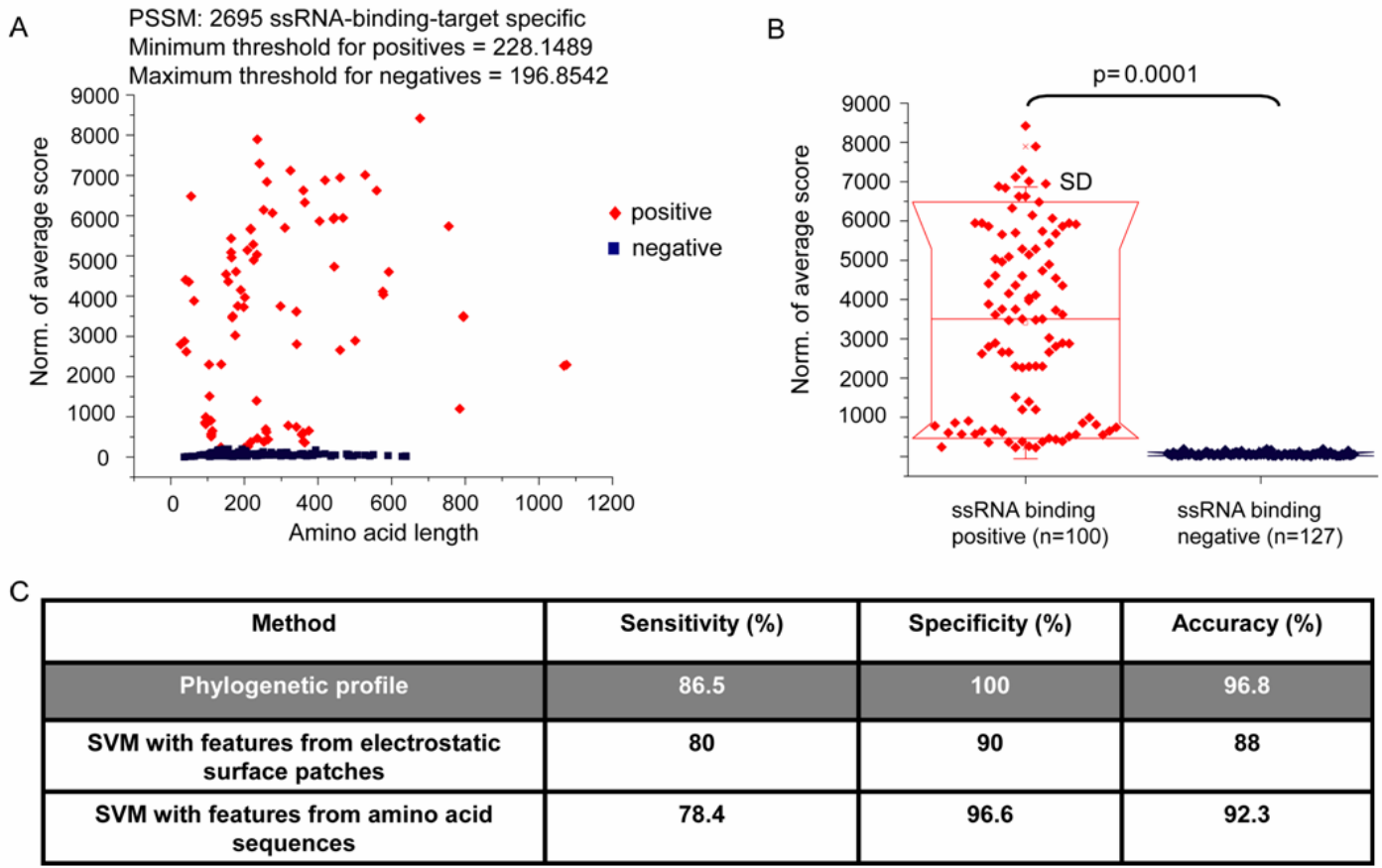

Figure 5. Classifying single-stranded RNA-binding proteins- (A) Norms of average scores for 100 positive (red) and 127 negative (black) sequences when analyzed using 2695 PSSMs specific to single-stranded binding domains of varying folds. (B) Box-plots with confidence values for the data in A. (C) Comparison of phylogenetic profiles with 2 support-vector machine learning methods for single-stranded 37 positive and 118 negative RNA binding proteins in an independent set. Testing and training sequences, as well as the ssRNA-binding PSSMs used are provided upon request.

Using this expanded functional library (2695 PSSMs), we then analyzed a training set containing 100 single-stranded RBPs and 127 negative sequences (Figure 5A-B). Under these conditions, we see a clear separation of positive and negative sequences. In our testing dataset, which is comprised of 37 positive and 118 negative sequences from the Shazman et al study (31), we achieve 100% specificity, 96.8% accuracy, and 86.5% sensitivity. In comparison, these authors reported 90% specificity, 88% accuracy, and 80% sensitivity (Figure 5C). Thus, with the expanded PSSM library, our results rival those previously obtained.

### A

| RBP Group | Method | TP | FN | Sensitivity (%) |
|---|---|---|---|---|
| dsDNA binding | Phylogenetic profile | 26 | 8 | 76.47 |
| | SVM | 19 | 15 | 55.88 |
| ssDNA binding | Phylogenetic profile | 16 | 1 | 94.12 |
| | SVM | 10 | 7 | 76.92 |
| dsRNA Binding | Phylogenetic profile | 20 | 0 | 100 |
| | SVM | 9 | 11 | 45 |
| ssRNA binding | Phylogenetic profile | 32 | 5 | 86.47 |
| | SVM | 29 | 8 | 78.38 |

### B

| Group | PSSM type | Sensitivity(%) | Specificity(%) | Accuracy (%) |
|---|---|---|---|---|
| dsRNA vs. dsDNA | dsRNA binding | 100 | 100 | 100 |
| | dsDNA binding | 76.47 | 85 | 79.6 |
| ssRNA vs. dsDNA | ssRNA binding | 86.49 | 97.06 | 91.55 |
| | dsDNA binding | 76.47 | 83.78 | 80.28 |
| ssRNA vs. ssDNA | ssRNA binding | 86.49 | 94.12 | 88.89 |
| | ssDNA binding | 94.12 | 94.59 | 94.44 |

**Figure 6. Classification of nucleotide-binding proteins- (A)** Table comparing the results between phylogenetic profiles and SVM (Shazman et al [2]) for identifying ssRNA (37 sequences), dsRNA (20 sequences), ssDNA (17 sequences), and dsDNA (34 sequences) nucleotide-binding (for sequences used and PSSMs in each library are available upon request). We observe that phylogenetic profiles outperform SVM for the detection of all classes of nucleotide-binding proteins. **(B)** Table displays the comparative accuracy of phylogenetic profiles for mixed-function/fold datasets (raw data is plotted in Supplemental Fig 4). In all cases, PSSM libraries trained to a particular nucleotide-binding property provide specific measures for their respective group (e.g. ssRNA vs. dsDNA, dsRNA vs. dsDNA, etc).

Under the same paradigm, we generated additional PSSMs for double-stranded RNA-binding (dsRNA), single-stranded DNA-binding (ssDNA), and double-stranded DNA binding (dsDNA) (Figure 6). We then compared our results to those from the Shazman et al study (31) for the classification of ssRNA and dsDNA-binding domains. These authors obtained 50% specificity, 51% accuracy, and 53% sensitivity within this dataset. Our results using either the ssRNA or dsDNA PSSM libraries are much improved: 97%/83% specificity, 91%/80% accuracy, and 86%/76% sensitivity, respectively. We also compared our results in additional testing sets curated from PROSITE (32), for proper classification of dsRNA vs. dsDNA, ssRNA vs. ssDNA, and ssRNA vs. dsDNA (Figure 6B). Although attempted, Shazman et al concluded that further refinement of their method was needed to accomplish these more difficult comparisons (31). Conversely, we obtain robust measurements for these comparisons (Figure 6), in particular for dsRNA binding, where we achieve 100% accuracy.

As a final comparison to the Shazman et al study (31), we investigated whether phylogenetic profiles could accurately discriminate between RRM domains that bind RNA and those that don't. Using electrostatic patches, they were able to accurately segregate these types of RRM domains with 100% accuracy. In our first attempt (Figure 7A), we could not segregate these RRM domains using our ssRNA PSSM library where RRM profiles were removed. To overcome this limitation, we designed a new residue-based phylogenetic profile (Figure 7B). This phylogenetic profile employs a residue distribution matrix developed by Bock et al (33). The matrix is composed of the frequency of the 20 amino acids comprising the protein, as well as 3 descriptors of chemical features. The 3 descriptors represent global composition of specific chemical groups, and consist of composition, transition, and distribution. The composition (C) is the number of amino acids within a particular group divided by total number of amino acid in all chemical groups. Transition (T) is the frequency of transition from one chemical group to another chemical group in a sequence. Distribution (D) is the chain length within the first, 25%, 50%, 75% and 100% of the amino acid in a specific chemical group (33). Following, we cluster the sequences using Hierarchical clustering and Pearson's correlation. The results from this experiment are presented in Figure 7C. Using this method, we observe a clear segregation of RNA-binding RRMs from non RNA-binding RRMs with excellent statistical support.

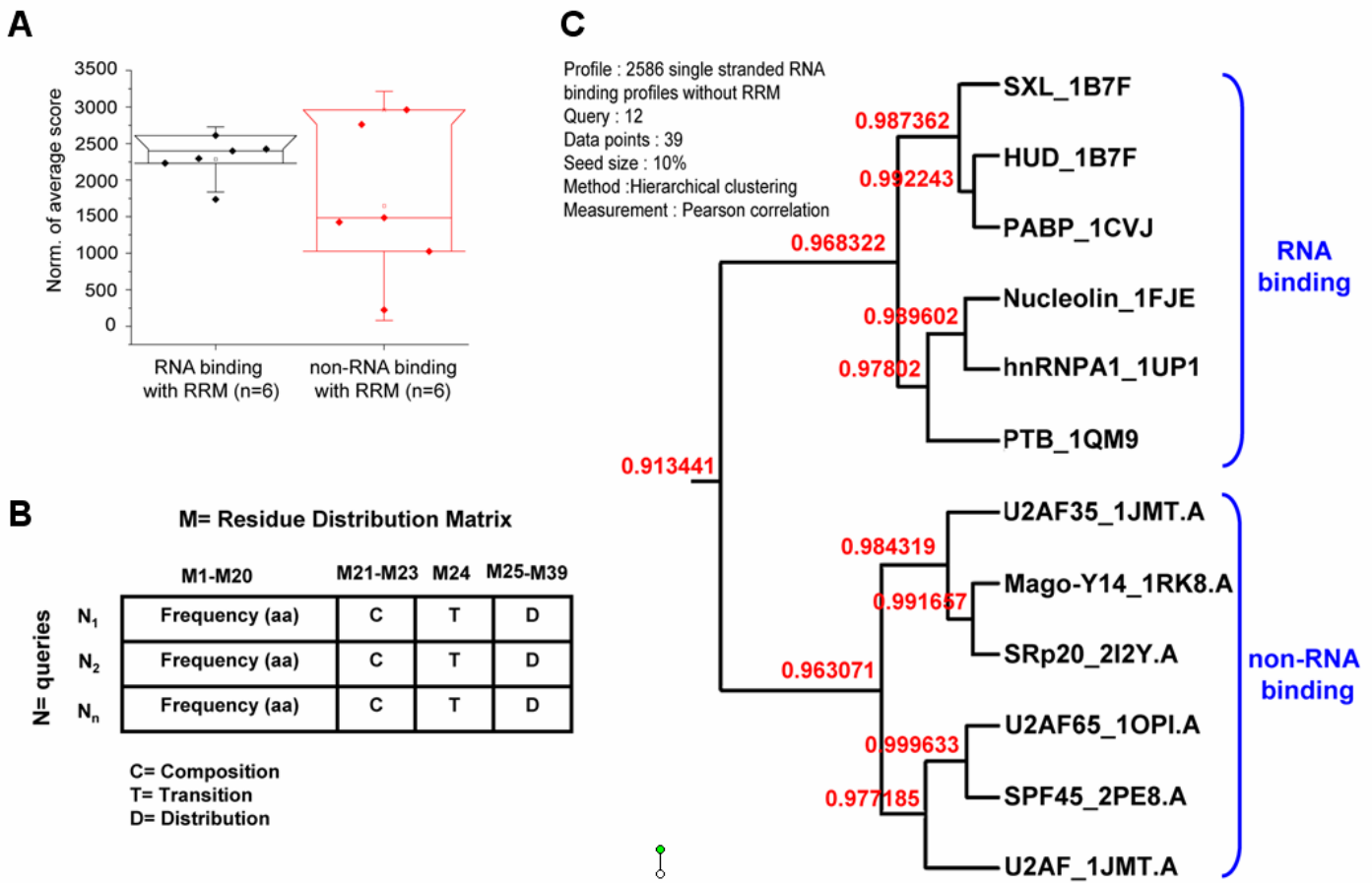

Figure 7. Classifying RRM domains with and without RNA-binding- (A) Box-plots of average score norms using 2695 ssRNA-binding PSSMs for RRM domains that are positive (red) or negative (black) for RNA-binding. We observe no significant difference between these two groups using this measurement. (B) *Residue Distribution Matrix:* The matrix consists of 20 amino acid compositions and 3 descriptors of chemical features. (C) Hierarchical clustering (Pearson's correlation) of the RRM domains from (A) after generating a residue distribution matrix for each sequence. We observe that in this analysis RNA binding and non-RNA binding RRM domains separate into two clades, each of which has excellent statistical support.

**Theories on Structural Homology Detection**

It has been proposed that the number of distinct native state folds is extremely limited(34). This suggests that with accurate measurements of homology, inferences of structure from primary amino acid sequences are possible. Indeed, significant advances have been made in this arena with tools such as Rosetta, PHYRE, and MUSTANG(35-37). In addition, algorithms which create structural predictions using physical laws and constraints also exist (e.g. ZAM, M-TASSER)(38;39). Although effective, all of these algorithms suffer from either computational constraints in predicting secondary structure (38;39) and/or barriers in homology detection(40;41). We theorize that a computational pipeline could be developed using phylogenetic profiles for rapid homology detection and secondary structure annotation. Once derived, this information could inform 3D prediction algorithms to improve their performance. To this end, we have initiated multiple lines of research to identify structural homology using benchmark structural datasets such SABmark, Balibase, and SCOP(42-44).

*Homology Detection*

As shown in Figure 8, phylogenetic profiles are capable of identifying homologous folds in sequences containing less than 25% identity. In this performance evaluation, we analyzed 534 proteins representing 61 SCOP folds from the Sabmark "twilight" dataset (42) with phylogenetic profiles. These results were compared to other benchmark methods using Receiver Operating Characteristics curve analysis. ROC curve shows the sensitivity (i.e. True Positive Pairs(TP)/(TP+False Negative

Pairs(FN)) of each method at different false positive rates (i.e. 1 – TN/(TN+False Positive Pairs)). As the ROC curve of a given method is closer to left-top corner of the graph, the method is considered more accurate. To calculate sensitivity at different false positive rate, given a query, top-K sequences with highest correlation or smallest e-value/p-value are considered to be returned as related by each method, as increasing K (see Methods for complete description).

Our results are very promising. Our preliminary data demonstrates that phylogenetic profiles generated using embedded alignment data generated using GDDA-BLAST (1) outperform all algorithms tested (Figure 8). These algorithms include SAMT2K, Prof Sim, and FFAS, with the latter considered to be the current "gold-standard" algorithm(45-47). Our performance on this dataset one year ago seemed to be at the "glass ceiling" as FFAS (red), GDDA-BLAST (magenta), SAMT2K (not shown), and Prof Sim (not shown) all had relatively comparable performance in this dataset (29%, 27%, 23%, and 20% sensitivity at false-positive rate<0.01, respectively).

We speculated that our performance could be improved by either optimizing our PSSM libraries and/or modification of our scoring schemes. To address the former, we created fold-specific libraries using representative sequences for each fold that do not appear in our testing set. After collecting representative sequences, we generated PSSMs from these sequences or their homologues using PSI-BLAST(17). Homologues of representative sequences were obtained using PSI-BLAST and the NCBI NR database. Each sequence is searched in the NCBI NR database using PSI-BLAST (e=1X10-6, 6 iterations). The homologous sequences obtained from this search are then converted PSSMs using PSI-BLAST. Following, redundancy is removed. Using this process, we obtained a total of 36,146 PSSMs for the 61 folds in our testing set. To test whether fold-specific PSSMs have improved utility, we altered how we encoded them into the phylogenetic profile. Specifically, using our previous multi-fold PSSM libraries, each PSSM represented a unique M for each query N. For our fold-specific PSSM libraries, each M represents the total score for the alignments obtained for each PSSM library. Thus for this experiment, the M dimension is equal to the number of unique fold libraries (M=61).

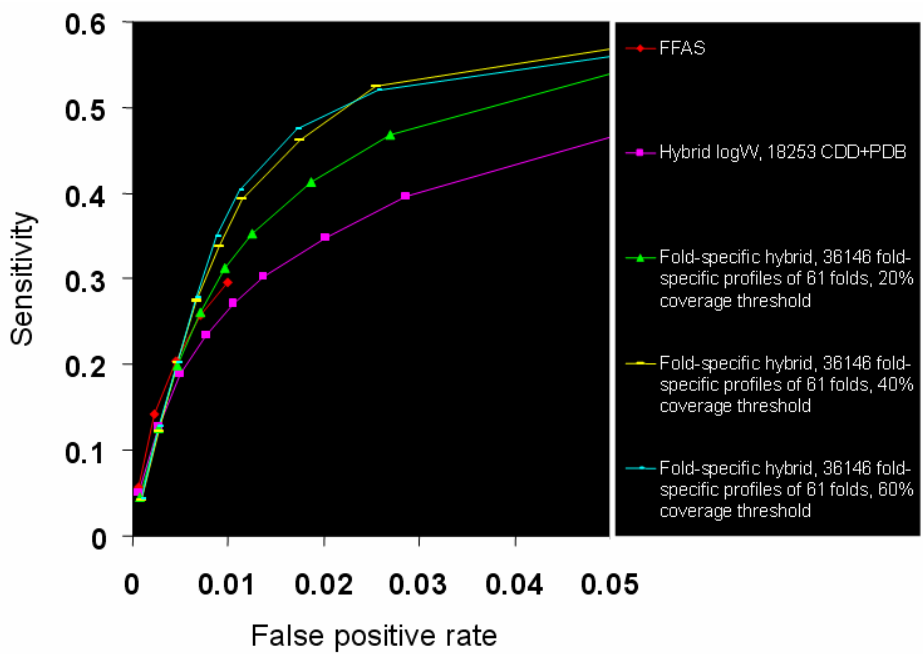

**Figure 8: ROC Curves on the "Twilight Zone" SABMARK Benchmark Dataset.** This dataset contains 534 queries with twilight zone sequence similarity representing 61 SCOP fold specific groups. For phylogenetic profiles, all protein pairs are encoded as a vector of composite scores (norm. # of hits, norm. %identity, norm. % coverage from 36,146 fold specific PSSMs) then each of the sequences is compared in an all-against-all fashion. To compare encoded sequences, Pearson's correlation coefficient is computed to measure the similarity between two sequences. All values for the Pearson's correlation are accepted into the comparative analysis. ROC performance analyses were performed by top-K analysis (Bradley, 1995) with largest correlation coefficient or smallest e-value. This sequence is considered homologous sequences to the query. If a sequence pair is actually related/homologous (in the same group) to the query, it is a termed a true positive. If a sequence is in the K sequences and not related to the query, it is a false positive.

In this study, we also tested different scoring schemes/conditions for GDDA-BLAST. The first is a hybrid log-weighted Pearson's correlation value. In this *Log Weighted* scoring scheme, three phylogenetic profiles are generated for each sequence with three different scores, such as *hit*, *coverage*, and *identity*, separately. Since a PSSM is likely to have no significant information to measure sequences if the profile is positive with many sequences, each *hit*, *coverage*, and *identity* score for each profile in a phylogenetic profile is adjusted by the frequency that the profile is positive on 534 sequences so that the scores of the profile has less weight when the correlation between two sequences is calculated. For example, the adjusted *hit* score of the original *hit* score *h* between a query and a profile *x* becomes $h \times \log_2(\frac{1}{\alpha/N})$ where $\alpha$ is # of sequences which the profile *x* is positive and N is the total number of sequences to compare (N=534 in our test set). Pearson's correlation between two sequences is calculated as multiplying Pearson's correlation of each different score of phylogenetic profiles of the sequences.

In our fold-specific hybrid scoring scheme, we calculate hits, %coverage, and %identity for each fold group by summing up each hits, %coverage, identity for each PSSM in the same fold group, then divide each sum by # of positive PSSMs in the group. Following, as for the Log Weighted scoring scheme, Pearson's correlation between two queries is calculated by multiplying Pearson's correlation of each different score of phylogenetic profiles of the queries.

The improvements to our PSSM libraries and scoring schemes were successful for improving our performance. Our best results are achieved using the Hybrid Log Weighted scoring scheme with a 60% coverage threshold (Figure 8 cyan), although this is negligibly better than a 40% coverage threshold (Figure 8 yellow). Importantly, our performance is ~11% better than FFAS (29% accuracy vs. GDDA-BLAST 40% sensitivity at false-positive rate of 0.01).

**Conclusions**

Although simple to ask, questions related to protein structure/function are often difficult to answer. Nevertheless, if these parameters could be accurately measured, our understanding of biological systems would be dramatically enhanced. One could envision a future where angstrom level structural models, comprehensive functional annotation, and precise evolutionary rates of proteins can all be achieved computationally from a single dataset. In this study, we determined that phylogenetic profiles provide a robust quantitative measurement of RNA-binding function as well as homology detection in the "twilight zone" of sequence similarity. These conclusions are supported by a number of experiments that demonstrate our ability to: (i) detect highly-divergent RBPs, (ii) discriminate between closely related targets (e.g. dsRNA vs. dsDNA), (iii) differentiate between classical RBP folds that bind RNA and those that don't, and (iv) outperform existing algorithms on the Sabmark "twilight-zone" dataset. Moreover, all of these results were obtained using only primary amino acid sequences, and in some cases their chemical features.

In our previous lipid-binding studies using phylogenetic profiles, we did not have numerical thresholds which could be used to threshold positive and negative results (1;13). Thus, these results were all confirmed experimentally (1;10;12;14;15). We chose to develop our quantitative thresholds using nucleotide binding as a benchmark function as they are comprised of well characterized yet diverse structural folds having a variety of ligand specificity (24;31;48). Our results demonstrate that construction of PSSMs that represent specific folds and/or activities can be used to generate accurate and sensitive phylogenetic profiles. Indeed, fold-specific PSSMs libraries can accurately identify homologous folds (Figure 3 and 8). Similarly, function-specific PSSMs libraries can accurately identify homologous functions in proteins of diverse structure (Figure 6). In addition, when these phylogenetic profiles are informed by amino acid physio-chemical properties (33), they are capable of differentiating activity within a specific fold (e.g. RRM) (Figure 7B-C).

Similar to accurately assigning nucleotide-binding function, identifying related structures of high divergence is also challenging with current technology. Indeed, there appears to be a "glass ceiling" in benchmark datasets, such as the Sabmark, whereby the most sensitive algorithms have only 25-30% sensitivity at statistical limits (false positive rate =0.01). Using our current measurements, phylogenetic profiles have started to break through the "glass ceiling" (~40% sensitivity at statistical limits). It is tantalizing to consider that optimization of our scoring schemes and PSSM libraries could detect even more distantly related structural homologues. Were this possible, approaches such as ours could inform state-of-the-art tertiary modeling programs. These programs often cannot perform structural modeling due, in large part, to rate-limiting calculations of secondary structure and/or lack of sensitivity for detecting homologous folds with high sequence divergence.

An implication that is emerging from our studies is that PSSMs derived from structurally distinct but functionally conserved sequences contain "functional homology" that can be measured. This functional homology may be due to the energetic properties necessary to support various protein-ligand interactions. Indeed, studies from Chen and Lim (49) demonstrate that the energetic features of DNA- and RNA-binding sites are distinct from each other, yet still maintain some global similarities. In our lipid-binding studies, our PSSM library was comprised of numerous folds for a variety of lipid-specificities (e.g. ENTH, C2, PH, FYVE, etc) (1). While diverse, these PSSMs have a propensity to align in similar regions which bind lipids as demonstrated by us and others (10;50-52). Similarly, PSSM libraries constructed from structurally diverse RBDs are specific for variety RNA-binding functions (e.g. mRNA, tRNA, snRNA, etc) (Figure 4). These results imply that this is a general phenomenon and suggests that there are conserved functional motifs that can exist in multiple structural contexts.

Since PSSM libraries can quantitatively relate divergent structures, it is reasonable to consider that these same libraries could identify novel RBDs in folds which have yet to be characterized. An analogous and relevant example is the TRP_2 domain, a conserved domain contained in all canonical transient receptor potential channels (TRPCs). We demonstrated that this domain is detected by lipid-binding PSSMs which do not contain this fold, and that TRP_2 domains bind lipid (10). Further, upon identification of novel folds, they can be incorporated into the PSSM library, iteratively improving it.

Should this be true, a powerful PSSM library could be generated for any function with a sufficient knowledgebase (e.g. kinase, phosphatase, ATP-binding, GTP-binding, etc); however, a single domain of known function is sufficient to create a sensitive PSSM library (e.g. KH, RRM, zinc-finger, etc). This approach could be used to annotate the vast numbers of DUF-containing proteins (domains of unknown function) and sequences containing no annotation. The fold-specific results in this manuscript and our ongoing studies strongly suggest that fold-function specific PSSMs provide the most sensitive/specific libraries (see Hong et al Physics Archives, Noveber 2009). Thus, as new structures/functions are identified, fold-function specific PSSM libraries can be generated using PSI-BLAST (17).

In conclusion, we propose that future work aimed at creating comprehensive and refined PSSM libraries. In order to accomplish this goal requires taking control of the established PSSM databases (e.g. CDD, Pfam, SMART, etc) by re-annotating them with standardized and experimentally validated ontology. Further, refining our methods for generating PSSM libraries has the potential to exponentially increase the structural/functional annotation of all classes of proteins across taxa. Such an advance would have broad impacts on human health and disease, as well as basic science endeavors.

**Acknowledgments**
This work was supported by the Searle Young Investigators Award and start-up money from PSU (RLP), NCSA grant TG-MCB070027N (RLP, DVR), The National Science Foundation 428-15 691M (RLP, DVR), and The National Institutes of Health R01 GM087410-01 (RLP, DVR). This project was also funded by a Fellowship from the Eberly College of Sciences and the Huck Institutes of the Life Sciences (DVR) and a grant with the Pennsylvania Department of Health using Tobacco Settlement Funds (DVR). The Department of Health specifically disclaims responsibility for any analyses, interpretations, or conclusions. We would especially like to thank Jason Holmes and the CAC center for their superior support. We would also like to thank Drs. Robert E. Rothe, Jim White, Barbara Van Rossum, F. Stone Capture, Flattus McGhee van Vollmar for creative dialogue.

**Methods**

*Collection of master sequences and generation of PSSMs*

To generate PSSMs for functional or structural-specific folds, we first collected master sequences from the NCBI NR database and the literature. All sequences were either structurally resolved or were experimentally validated for the desired function. All PSSM master sequences used in this study are available upon request.

**Preparation of function or structure-specific PSSM set**

To generate the PSSM set for a specific protein function or structure fold, we first collected protein sequences which are known to be related to the function or structure of our interest. For the PSSM set, we generated PSSMs with the collected sequences or the sequences expanded from the collected sequences using PSI-BLAST. For expansion, each collected sequence is searched against NCBI NR database by PSI-BLAST (with the option of -e $1X10^{-3}$ –h $1X10^{-6}$). Among the returned sequences, we filtered out the sequences whose pairwise identity to the query sequence is more than 90% and redundant sequences in the set. And, for PSSM generation for those expanded sequences, PSI-BLAST (with the option of –h $1X10^{-6}$) was used again. All PSSM sets, which have been used for our test, will be provided upon request.

Datasets for our experiments contain sets of RBPs and non-nucleotide binding proteins (NNBPs). The sets of RBPs consist of specific (single-fold) RBP and target-based (multi-fold) RBP groups. Sets of single-fold RBPs were constructed using RNA binding motif as defined in Chen et al [3]. Based on these definitions, proteins were searched by key words (e.g. RNA-binding, DNA-binding, nucleotide-binding, etc) in NCBI and PROSITE protein databases, and manually verified for each class. Sets of multi-fold RBPs were built for mRNA-, tRNA-, rRNA-, and snRNA-binding proteins. Sequences for each class were collected by keyword searches and manually verified. The NNBPs dataset was obtained from Stawiski *et al.* [13]. The RBP and DBP datasets used in Figure 7 were obtained from Shazman et al [2]. All datasets and sequences are available upon request.

*Phylogenetic profile*
See Hong et al Physics Archives, November 2009 for complete description.

*Calculation of scores*

Theoretically, a protein should align best with PSSMs of similar fold and/or function. Based on this hypothesis, we have developed a scoring scheme to calculate a residue score which represents the occurrence of identical and positive residues from each query-PSSM alignment above threshold. Raw scores for each residue are calculated by scoring a value=2 for identities and value=1 for positive substitutions from each alignment. These values are then summed for all alignments at each position to obtain a total residue score. Following, these scores are normalized using the series of equations shown below. Equation (1) finds highly conserved residues whose score is above the average residue score in the sequence. Equation (2) recalculates the average score of these residues as a representative score for each sequence. Equation (3) calculates the norms of average scores to reduce the effect of the length. This data can be used to obtain thresholds with positive and negative training sets. The positive threshold is the minimum score from the positive group and the negative threshold is the maximal score from the negative group.

$$NS_{residue} = SC_{raw} - \left( SUM_{total\ score} / LEN_{query} \right) \quad (1) \qquad ASC_{query} = Sum_{PNS} / NR_{residue} \quad (2)$$

$$NORM_{average\ score} = AS_{query} * 100 / LEN_{query} * 2 \quad (3)$$

where $NS_{residue}$ = The normalized score of a residue,
$SC_{raw}$ = The raw score of a residue,
$SUM_{total\ score}$ = The sum of total query score,
$LEN_{query}$ = The query length,
$ASC_{query}$ = The average query score,
$SUM_{PNS}$ = The sum of positive normalized scores
$NR_{residue}$ = The number of residues with positive scores
$NORM_{average\ score}$ = The norm of an average score
$AS_{query}$ = The average query score

*Residue Distribution Matrix*

Using the amino acid positions as defined by GDDA-BLAST, we incorporated the residue distribution matrix developed by Bock et al (33). This matrix measures the chemical characteristics of a protein based on its primary amino acid sequence. The results obtained from this matrix are then included in our NXM phylogenetic profiles. Briefly, domain boundaries are calculated for each query from the overlapping PSSM alignments as previously described (1;15). Following, amino-acid positions are isolated using Equations 1-3 (see calculation of scores). Equation (4) calculates the amino acid frequency of all residues obtained from equation (3). We then calculate the composition (C), transition (T), and distribution (D) of amino acids in their physio-chemical groups. The composition (C) is the number of amino acids with a particular group (i.e. hydrophobic, positively and negatively charged residues) divided by total number of amino acid in all chemical groups. The transition (T) is the frequency of transition from one chemical group to another chemical group in the sequence. The distribution (D) is the chain length within the first, 25%, 50%, 75% and 100% of the amino acid for a specific chemical group [9]. Therefore, our matrix is composed of 20 features of the frequency, 3 composition features, 1 transition feature, and 15 distribution features. These matrices are then clustered in the N-dimension by hierarchical clustering using Pearson's correlation.

$$\text{The composition of an amino acid} = \frac{\#\ of\ an\ amino\ acid}{\#\ of\ total\ amino\ acids} \times 100 \quad (4)$$

*Statistical Analysis*

Using discriminative measurements [19,20], we assayed the performance of all algorithms used in the study for sensitivity, specificity, and accuracy. Abbreviations: true positives (TP), true negatives (TN), false positives (FP) and false negatives (FN). All p-values in this study were derived using Student T-test, with error bars reflecting the standard deviation of the dataset.

$$\text{Sensitivity} = \frac{TP}{TP+FN} \times 100\% \quad (5) \qquad \text{Specificity} = \frac{TN}{TN+FP} \times 100\% \quad (6)$$

$$\text{Accuracy} = \frac{TP+TN}{TP+TN+FP+FN} \times 100\% \quad (7)$$

### Reference List

1. Ko,K.D., Hong,Y., Chang,G.S., Bhardwaj,G., van Rossum,D.B., and Patterson,R.L. 2008. Phylogenetic Profiles as a Unified Framework for Measuring Protein Structure, Function and Evolution. *Physics Archives* **arXiv:0806.239, q-bio.Q**.

2. Ponting,C.P., and Russell,R.R. 2002. The natural history of protein domains. *Annu. Rev. Biophys. Biomol. Struct.* **31**:45-71.

3. Vogel,C., Bashton,M., Kerrison,N.D., Chothia,C., and Teichmann,S.A. 2004. Structure, function and evolution of multidomain proteins. *Curr. Opin. Struct. Biol.* **14**:208-216.

4. Park,J., Karplus,K., Barrett,C., Hughey,R., Haussler,D., Hubbard,T., and Chothia,C. 1998. Sequence comparisons using multiple sequences detect three times as many remote homologues as pairwise methods. *J. Mol. Biol.* **284**:1201-1210.

5. Lassmann,T., and Sonnhammer,E.L. 2005. Kalign--an accurate and fast multiple sequence alignment algorithm. *BMC. Bioinformatics.* **6**:298.

6. Pellegrini,M., Marcotte,E.M., Thompson,M.J., Eisenberg,D., and Yeates,T.O. 1999. Assigning protein functions by comparative genome analysis: protein phylogenetic profiles. *Proc. Natl. Acad Sci U. S. A* **96**:4285-4288.

7. Kim,Y., and Subramaniam,S. 2006. Locally defined protein phylogenetic profiles reveal previously missed protein interactions and functional relationships. *Proteins* **62**:1115-1124.

8. Wu,J., Mellor,J.C., and DeLisi,C. 2005. Deciphering protein network organization using phylogenetic profile groups. *Genome Inform.* **16**:142-149.

9. Chang,G.S., Hong,Y., Ko,K.D., Bhardwaj,G., Holmes,E.C., Patterson,R.L., and van Rossum,D.B. 2008. Phylogenetic profiles reveal evolutionary relationships within the "twilight zone" of sequence similarity. *Proc. Natl. Acad Sci U. S. A* **105**:13474-13479.

10. van Rossum,D.B., Oberdick,D., Rbaibi,Y., Bhardwaj,G., Barrow,R.K., Nikolaidis,N., Snyder,S.H., Kiselyov,K., and Patterson,R.L. 2008. TRP_2, a Lipid/Trafficking Domain That Mediates Diacylglycerol-induced Vesicle Fusion. *J. Biol. Chem.* **283**:34384-34392.

11. Chakraborty,A., Koldobskiy,M.A., Sixt,K.M., Juluri,K.R., Mustafa,A.K., Snowman,A.M., van Rossum,D.B., Patterson,R.L., and Snyder,S.H. 2008. HSP90 regulates cell survival via inositol hexakisphosphate kinase-2. *Proc. Natl. Acad Sci U. S. A* **105**:1134-1139.

12. Mustafa,A.K., van Rossum,D.B., Patterson,R.L., Maag,D., Ehmsen,J.T., Gazi,S.K., Chakraborty,A., Barrow,R.K., Amzel,L.M., and Snyder,S.H. 2009. Glutamatergic regulation of serine racemase via reversal of PIP2 inhibition. *Proc. Natl. Acad. Sci. U. S. A*.

13. van Rossum,D.B., Patterson,R.L., Sharma,S., Barrow,R.K., Kornberg,M., Gill,D.L., and Snyder,S.H. 2005. Phospholipase Cgamma1 controls surface expression of TRPC3 through an intermolecular PH domain. *Nature* **434**:99-104.

14. Caraveo,G., van Rossum,D.B., Patterson,R.L., Snyder,S.H., and Desiderio,S. 2006. Action of TFII-I outside the nucleus as an inhibitor of agonist-induced calcium entry. *Science* **314**:122-125.

15. Hong,Y., Chalkia,D., Ko,K.D., Bhardwaj,G., Chang,G.S., van Rossum,D.B., and Patterson,R.L. 2009. Phylogenetic Profiles Reveal Structural and Functional Determinants of Lipid-binding. *Journal of Proteomics and Bioinformatics* 139-149.

16. Zachos,N.C., van Rossum,D.B., Li,X., Caraveo,G., Sarker,R., Cha,B., Mohan,S., Desiderio,S., Patterson,R.L., and Donowitz,M. 2009. Phospholipase C-{gamma} binds directly to the Na+/H+ exchanger 3 and is required for calcium regulation of exchange activity. *J. Biol. Chem.*

17. Altschul,S.F., Madden,T.L., Schaffer,A.A., Zhang,J., Zhang,Z., Miller,W., and Lipman,D.J. 1997. Gapped BLAST and PSI-BLAST: a new generation of protein database search programs. *Nucleic Acids Res.* **25**:3389-3402.

18. Schaffer,A.A., Wolf,Y.I., Ponting,C.P., Koonin,E.V., Aravind,L., and Altschul,S.F. 1999. IMPALA: matching a protein sequence against a collection of PSI-BLAST-constructed position-specific score matrices. *Bioinformatics* **15**:1000-1011.

19. Henikoff,S., and Henikoff,J.G. 1997. Embedding strategies for effective use of information from multiple sequence alignments. *Protein Sci.* **6**:698-705.

20. Sonnhammer,E.L., Eddy,S.R., and Durbin,R. 1997. Pfam: a comprehensive database of protein domain families based on seed alignments. *Proteins* **28**:405-420.

21. Letunic,I., Copley,R.R., Schmidt,S., Ciccarelli,F.D., Doerks,T., Schultz,J., Ponting,C.P., and Bork,P. 2004. SMART 4.0: towards genomic data integration. *Nucleic Acids Res.* **32 Database issue**:D142-D144.

22. Marchler-Bauer,A., Anderson,J.B., Cherukuri,P.F., Weese-Scott,C., Geer,L.Y., Gwadz,M., He,S., Hurwitz,D.I., Jackson,J.D., Ke,Z. et al 2005. CDD: a Conserved Domain Database for protein classification. *Nucleic Acids Res.* **33 Database Issue**:D192-D196.

23. Mulder,N., and Apweiler,R. 2007. InterPro and InterProScan: tools for protein sequence classification and comparison. *Methods Mol. Biol.* **396**:59-70.

24. Chen,Y., and Varani,G. 2005. Protein families and RNA recognition. *FEBS J.* **272**:2088-2097.

25. Karplus,K., Sjolander,K., Barrett,C., Cline,M., Haussler,D., Hughey,R., Holm,L., and Sander,C. 1997. Predicting protein structure using hidden Markov models. *Proteins* **Suppl 1**:134-139.

26. Han,L.Y., Cai,C.Z., Lo,S.L., Chung,M.C., and Chen,Y.Z. 2004. Prediction of RNA-binding proteins from primary sequence by a support vector machine approach. *RNA.* **10**:355-368.

27. Shamoo,Y., bdul-Manan,N., and Williams,K.R. 1995. Multiple RNA binding domains (RBDs) just don't add up. *Nucleic Acids Res.* **23**:725-728.

28. Corsini,L., Bonnal,S., Basquin,J., Hothorn,M., Scheffzek,K., Valcarcel,J., and Sattler,M. 2007. U2AF-homology motif interactions are required for alternative splicing regulation by SPF45. *Nat. Struct. Mol. Biol.* **14**:620-629.

29. Stawiski,E.W., Gregoret,L.M., and Mandel-Gutfreund,Y. 2003. Annotating nucleic acid-binding function based on protein structure. *J. Mol. Biol.* **326**:1065-1079.

30. Cai,C.Z., Han,L.Y., Ji,Z.L., Chen,X., and Chen,Y.Z. 2003. SVM-Prot: Web-based support vector machine software for functional classification of a protein from its primary sequence. *Nucleic Acids Res.* **31**:3692-3697.

31. Shazman,S., and Mandel-Gutfreund,Y. 2008. Classifying RNA-binding proteins based on electrostatic properties. *PLoS. Comput. Biol.* **4**:e1000146.

32. Bairoch,A. 1991. PROSITE: a dictionary of sites and patterns in proteins. *Nucleic Acids Res.* **19 Suppl**:2241-2245.

33. Bock,J.R., and Gough,D.A. 2001. Predicting protein--protein interactions from primary structure. *Bioinformatics* **17**:455-460.

34. Chothia,C. 1992. Proteins. One thousand families for the molecular biologist. *Nature* **357**:543-544.

35. Simons,K.T., Bonneau,R., Ruczinski,I., and Baker,D. 1999. Ab initio protein structure prediction of CASP III targets using ROSETTA. *Proteins* **Suppl 3**:171-176.

36. Konagurthu,A.S., Whisstock,J.C., Stuckey,P.J., and Lesk,A.M. 2006. MUSTANG: a multiple structural alignment algorithm. *Proteins* **64**:559-574.

37. nett-Lovsey,R.M., Herbert,A.D., Sternberg,M.J., and Kelley,L.A. 2008. Exploring the extremes of sequence/structure space with ensemble fold recognition in the program Phyre. *Proteins* **70**:611-625.

38. Shell,M.S., Ozkan,S.B., Voelz,V., Wu,G.A., and Dill,K.A. 2009. Blind test of physics-based prediction of protein structures. *Biophys. J.* **96**:917-924.

39. Chen,H., and Skolnick,J. 2008. M-TASSER: an algorithm for protein quaternary structure prediction. *Biophys. J.* **94**:918-928.

40. Blake,J.D., and Cohen,F.E. 2001. Pairwise sequence alignment below the twilight zone. *J. Mol. Biol.* **307**:721-735.

41. Yona,G., and Levitt,M. 2002. Within the twilight zone: a sensitive profile-profile comparison tool based on information theory. *J. Mol. Biol.* **315**:1257-1275.

42. Van,W., I, Lasters,I., and Wyns,L. 2005. SABmark--a benchmark for sequence alignment that covers the entire known fold space. *Bioinformatics.* **21**:1267-1268.

43. Thompson,J.D., Plewniak,F., and Poch,O. 1999. BAliBASE: a benchmark alignment database for the evaluation of multiple alignment programs. *Bioinformatics.* **15**:87-88.

44. Murzin,A.G., Brenner,S.E., Hubbard,T., and Chothia,C. 1995. SCOP: a structural classification of proteins database for the investigation of sequences and structures. *J. Mol. Biol.* **247**:536-540.

45. Karplus,K., Katzman,S., Shackleford,G., Koeva,M., Draper,J., Barnes,B., Soriano,M., and Hughey,R. 2005. SAM-T04: what is new in protein-structure prediction for CASP6. *Proteins* **61 Suppl 7**:135-142.

46. Jaroszewski,L., Rychlewski,L., Li,Z., Li,W., and Godzik,A. 2005. FFAS03: a server for profile--profile sequence alignments. *Nucleic Acids Res.* **33**:W284-W288.

47. Mittelman,D., Sadreyev,R., and Grishin,N. 2003. Probabilistic scoring measures for profile-profile comparison yield more accurate short seed alignments. *Bioinformatics* **19**:1531-1539.

48. Chen,Y.C., and Lim,C. 2008. Predicting RNA-binding sites from the protein structure based on electrostatics, evolution and geometry. *Nucleic Acids Res.* **36**:e29.

49. Chen,Y.C., and Lim,C. 2008. Common physical basis of macromolecule-binding sites in proteins. *Nucleic Acids Res.* **36**:7078-7087.

50. Kwon,Y., Hofmann,T., and Montell,C. 2007. Integration of phosphoinositide- and calmodulin-mediated regulation of TRPC6. *Mol. Cell* **25**:491-503.

51. Nilius,B., Mahieu,F., Prenen,J., Janssens,A., Owsianik,G., Vennekens,R., and Voets,T. 2006. The Ca2+-activated cation channel TRPM4 is regulated by phosphatidylinositol 4,5-biphosphate. *EMBO J.* **25**:467-478.

52. Zhang,H., Craciun,L.C., Mirshahi,T., Rohacs,T., Lopes,C.M., Jin,T., and Logothetis,D.E. 2003. PIP(2) activates KCNQ channels, and its hydrolysis underlies receptor-mediated inhibition of M currents. *Neuron* **37**:963-975.